\documentclass[%
twocolumn,
groupedaddress,
amsmath,amssymb,
aps,
pra,
]{revtex4-2}

\usepackage{graphicx}
\usepackage{amsfonts}
\usepackage{bm}
\usepackage{hyperref,url}
\usepackage{changes}
\usepackage{xcolor}
\usepackage{soul}

\newcommand{\bd}{\begin{displaymath}}
\newcommand{\ed}{\end{displaymath}}
\newcommand{\be}{\begin{equation}}
\newcommand{\ee}{\end{equation}}
\newcommand{\ba}{\begin{eqnarray}}
\newcommand{\ea}{\end{eqnarray}}

\begin{document}

\title[]{Partially coherent Airy beams: A cross-spectral density approach}

\author{R. Mart{\'\i}nez-Herrero}
\email{r.m-h@fis.ucm.es}
\affiliation{Department of Optics, Faculty of Physical Sciences,
Universidad Complutense de Madrid,\\
Pza.\ Ciencias 1, Ciudad Universitaria E-28040 Madrid, Spain}

\author{A. S. Sanz}
\email{a.s.sanz@fis.ucm.es}
\affiliation{Department of Optics, Faculty of Physical Sciences,
Universidad Complutense de Madrid,\\
Pza.\ Ciencias 1, Ciudad Universitaria E-28040 Madrid, Spain}

\begin{abstract}
Airy beams are known for displaying shape invariance and self-acceleration
along the transverse direction while they propagate forwards.
Although these properties could be associated with the beam coherence, it has been
revealed that they also manifest in the case of partially coherent Airy-type beams.
Here, these properties are further investigated by introducing and analyzing a class
of partially coherent Airy beams under both infinite and finite energy conditions.
The key element within the present approach is the so-called cross-spectral density, which
enables a direct connection with the quantum density matrix, making the analysis exportable
to the quantum realm to study the dynamics of Airy wave packets acted by both incoherence
and decoherence.
As it is shown, in the case of infinite energy beams both properties are preserved even under
the circumstance of total incoherence provided the underlying structure of the beam remains
equal to that of an Airy beam.
In the case of finite energy beams, a situation closer to a realistic scenario, as
experimental beams cannot have an infinite extension, it is shown that a propagation
range along which both properties are preserved can be warranted.
This is controlled by a critical distance, which depends on the spread range determined by
the parameters ruling the extension of random field spatial fluctuations.
Such a distance is determined by defining a position-dependent parameter that quantifies the
degree of overlapping between the propagated beam and the input one displaced by an amount
equivalent to the propagation distance.
\end{abstract}




\maketitle


\section{Introduction}
\label{sec1}

Airy wave packets are known for exhibiting two intriguing and counter
intuitive properties: when they are freely released in space, their
propagation is dispersionless and uniformly accelerated.
This self-accelerating solution to the free-particle nonrelativistic
Schr\"odinger equation was first noticed by Berry and Balazs in the
late 1970s \cite{berry:AJP:1979}, who provided a semiclassical
explanation to this behavior based on the concept of caustic.
Yet, by invoking the equivalence principle, it is also possible
an alternative explanation, as it was suggested shortly afterwards by
Greenberger \cite{greenberger:AJP:1980}.
Accordingly, the Airy wave packet can be understood as the stationary
solution for a free-falling system in a uniform gravitational field.
Later works have been aimed at finding first-principles derivations of
these puzzling solutions \cite{unnikrishnan:AJP:1996}, generalizing
the behavior to higher dimensions \cite{besieris:AJP:1994}, or even
exploring ways to redirect electron Airy beams without causing any
dispersion by means of magnetic fields \cite{efremidis:PRA:2021}.

Matter-wave Airy-type beams and their dynamics have also received attention
in the context of nonlinear extensions of the Schr\"odinger equation.
This is the case, for instance, of the Gross-Pitaevskii equation, a nonlinear
Schr\"odinger equation accounting for the dynamics of Bose-Einstein condensates
(BECs) in the mean-field (Hartree-Fock) approximation \cite{pethick2008bose}.
Within this context, it has been shown that, by properly controlling the
amplitude and phase of the BEC, it is possible to make it to behave as an
ordinary self-accelerating Airy beam, and also to produce an abrupt
autofocusing \cite{efremidis:PRA:2013}.
Airy-type solutions for BECs confined within time-dependent harmonic traps
have also been determined \cite{yuce:MPLB:2015}, which are shown to
spontaneously break the parity and time-reversal symmetries.
On the other hand, within the classical realm, experiments reported in the
literature show the emergence of Airy-type solutions in surface gravity
water waves \cite{arie:PRL:2015-1,arie:Fluids:2019}, where the behavior of
such waves is ruled by equations isomorphic to Schr\"odinger's one in the
linear regime.

In spite of all the literature generated around Airy (matter-)wave packets
at, say, a fundamental level, covering a wide variety of formal aspects
and physical situations, it is worth noting how experimentally-oriented
works involving matter waves are rather scarce instead.
The first experimental realization and observation of free-electron Airy
wave packets were reported by Voloch-Bloch {\it et al.}~\cite{voloch:Nature:2013}.
These Airy wave packets were produced by diffracting electron beams with
nanoscale holograms.
However, unlike Berry and Balazs formerly suggested, the beams
generated were not direct solutions of the nonrelativistic Schr\"odinger
equation, but paraxial solutions to the Klein-Gordon equation, which
suffices to describe the slowly changing transverse electron motion
in transmission electron microscopy.
Interestingly, these works can be considered to be within the subject
of electron beam shaping \cite{arie:PhysScr:2019}, aimed at producing
nanostructured electron beams \cite{arie:PRL:2015-2}, which is in direct
analogy to field of structured light beams, where paraxial treatments
are also common.

Now, it is precisely within the above-mentioned scenario of paraxiality and
structured light, where optical Airy beams have received much attention from
an experimental point of view since the mid-2000s \cite{christodoulides:Optica:2019}.
Note that translating the concept of Airy wave packet to the optical realm is
straightforward by virtue of the well-known isomorphism exhibited by the
nonrelativistic Schr\"odinger equation and the Helmholtz equation in paraxial
form, with the longitudinal coordinate in the latter playing the role of the
time in the former.
By producing the appropriate hologram phase mask, Siviloglou
{\it et al.}~\cite{christodoulides:PRL:2007} reported on the first experimental
evidence of light Airy beams (and of Airy beams, in general, since matter-wave
realizations came several years later).
In contrast to other types of structured light, Airy beams are characterized
for keeping their shape invariance (a transverse profile with the form of an Airy
function) while they propagate forward along the longitudinal coordinate, i.e.,
when they are observed at different output planes.
Moreover, the transverse evolution of the beam manifests the self-accelerating
motion of Airy wave packets, although this dependence is on the longitudinal
coordinate (output plane position).

By definition, light Airy beams correspond to an ideal situation in which the beam
has an infinite extension.
Hence, they involve an unbound (infinite) amount of energy.
To some extent, this behavior has been experimentally implemented and observed
\cite{christodoulides:PRL:2007}, but at the expense that, at some point (in space),
the beam produced deviates from the ideal Airy beam behavior; obviously, it is
not possible to implement a beam carrying an infinite amount of energy.
In this regard, since the first experimental demonstrations of Airy beams, the
interest has turned towards the so-called partially coherent Airy beams, which
correspond to more realistic physical implementations of also shape-invariant
self-accelerating light beams.

An appropriate tool to cope with the issue of partially coherent light is
the cross-spectral density (CSD) \cite{mandelwolf-bk}.
As it is known, apart from Airy beams, a significant class of sources is
characterized by keeping their shape invariant for paraxial propagation
conditions at any distance from the source, except for a transverse scaling
factor and a spherical curvature term.
This invariance concerns the whole CSD of the field and hence both the profile
displayed by the intensity distribution along the transverse direction and the
coherence properties.
When this condition is satisfied, the field CSD is said to be shape-invariant.
Research on possible forms of CSDs giving rise to fields with peculiar propagation properties
has been ongoing since the late 1970s (see, for instance, Ref.~\cite{korotkova:ProgOpt:2020},
and references therein).
Yet, there is a major inconvenience in devising new CSD forms, as they are not generic
two-point space functions.
A necessary and sufficient condition for a function to represent
a valid CSD is that it must be a non-negative definite Hermitian function
\cite{mandelwolf-bk}.
Whenever this condition is satisfied, the CSD is said to be genuine or bona-fide; if the
condition is not satisfied, then the function cannot represent the CSD of a possible
physical source.
In general, though, it is not easy to check the non-negativity of an integral kernel,
although genuineness criteria have been introduced in the literature with this purpose
\cite{gori:OptLett:2007,martinezherrero:OptLett:2009-1,martinezherrero:OptLett:2009-2,martinezherrero:JOSAA:2021}.

Here we approach and investigate partially coherent Airy-type beams from the point of
view of their corresponding CSDs in cases of both finite and infinite energy.
As it is shown, although the analysis in all cases starts from well-defined Airy
functions, the introduction of random fluctuations and, more specifically, the spread
functions that determine the spatial reach of such fluctuations are going to play a major
role in the preservation of the beam properties, namely, shape-invariance and
self-acceleration.
In the infinite energy case, we show that these properties still remain even when the fast
oscillations that characterize the decaying tail of Airy beams are smoothed out by the
averaging over random realizations.
This apparently counter intuitive behavior can be explained taking into account that a swarm
of identical Airy beams still underneath the internal structure of the full beam, regardless
of the extent of random fluctuations.
Since all single Airy beams included in the random mixture are independent one another,
their shape-invariance and self-acceleration properties remain unaffected, and hence
also those of the resulting partially coherent beam.
In the finite energy case, however, because additional constraints are set upon the
random displacements (leading to correlations among them), both shape-invariance and
self-acceleration can only be preserved up to a certain point.
This validity range is a function of the parameters governing the spread of the random
fluctuations, as it is shown by computing the overlapping between the propagated beam
and the translated initial beam.
Actually, by means of such calculations, it is shown that we can uniquely get an
estimate of the range of output planes where the beams are still going to be behave
in an Airy-type manner in simple terms.

According to the above discussion, the work is organized as follows.
The main aspects involved in the CSD approach here considered and its application
to the case of infinite energy are discussed in Sec.~\ref{sec2}.
Without loss of generality, this is illustrated by considering a Gaussian spread
function for the spatial random fluctuations acting on the beam at each position.
In Sec.~\ref{sec3}, two types of finite-energy, partially coherent Airy beams are
presented after generalizing the model of Sec.~\ref{sec2} with the introduction of
spatial correlations between the random displacements.
Although the starting point in the construction of each type of CSD is physically
different, eventually the functional form displayed by the general expression is shown
to be similar.
Finally, the main conclusions extracted from the work are summarized in Sec.~\ref{sec4}.


\section{Infinite Energy Beams}
\label{sec2}

For simplicity, the analysis will be constrained to a single transverse direction
instead of the full transverse plane, as we only need a transverse coordinate to
investigate the Airy-beam properties of shape-invariance and self-acceleration in partially
coherent beams.
Thus, to start with, consider a field amplitude that satisfies the paraxial Helmholtz
equation,
\be
 i\ \frac{\partial \mathcal{U} (x,z)}{\partial z} =
   - \frac{1}{2} \frac{\partial^2 \mathcal{U} (x,z)}{\partial x^2} ,
 \label{eq2}
\ee
where $x$ and $z$ denote, respectively, the transverse and longitudinal dimensionless
coordinates (they are referred to a certain characteristic length $\ell$ and the wave
number $k=2\pi/\lambda$, i.e., $x_{\rm dim} = \ell x$ and $z_{\rm dim} = k \ell^2 z$).
If a quasi-monochromatic beam consists of a number of random realizations of field
amplitudes $\mathcal{U}$, all satisfying Eq.~(\ref{eq2}), it can be specified by
its CSD,
\be
 \mathcal{W}(x,x',z) = \langle \mathcal{U}^*(x,z) \ \! \mathcal{U}(x',z) \rangle ,
 \label{eq1}
\ee
where the explicit dependence on the frequency is implicitly assumed.

The CSD (\ref{eq1}) provides us with a map of two-point field correlations at a given
output plane $z$, and hence with valuable information on the beam intensity distribution
and coherence properties.
Note that the beam intensity distribution is directly obtained from the diagonal of the
CSD (i.e., for $x' = x$) at a plane $z$, as
\be
 I(x,z) = \mathcal{W}(x,x,z) ,
 \label{eq3}
\ee
while the off-diagonal elements render information about the coherence between two
different points $x$ and $x'$.
Actually, there is a direct relationship between the CSD and quantifiers of the beam
coherence properties, such as the complex degree of coherence \cite{mandelwolf-bk},
\be
 \gamma (x,x',z) = \frac{\mathcal{W}(x,x',z)}{\sqrt{I(x,z)} \sqrt{I(x',z)}} .
 \label{eq4}
\ee
The modulus of this quantity ranges from 0, in the case of total incoherence, to a
maximum smaller or equal to 1, in the case of full coherence.
There are other related coherence quantifiers, such as the visibility \cite{mandelwolf-bk}
or the which-path distinguishability \cite{qureshi:OptLett:2021}, which additionally allow
us to establish a direct connection with matter waves, where the density operator plays the
role of the CSD.
In this regard, the discussion below can straightforwardly extended to the quantum realm.

Let us now consider that the field amplitude at the input plane $z=0$ is described by an
Airy function,
\ba
 Ai(x) & = & \frac{1}{\pi}\ \int_0^{\infty} \cos \left( \frac{u^3}{3} + x u \right) du
 \nonumber \\
 & = & \frac{1}{2\pi}\ \int_{-\infty}^{\infty} e^{i(u^3/3 + x u)} du .
  \label{v0}
\ea
This function satisfies the orthogonality condition
\be
 \int_{-\infty}^\infty Ai(x-u) Ai(x-v) \ \! dx = \delta (u-v) .
 \label{ortho}
\ee
From the input ansatz, Eq.~(\ref{v0}), we can obtain the field amplitude at any other
output plane $z \neq 0$, solution of the paraxial Helmholtz equation, Eq.~(\ref{eq2}), by
applying the free-space propagator.
This propagated solution reads
\be
 Ai(x,z) = e^{i(x - z^2/6)z/2} Ai(x - z^2/4) ,
 \label{evolvz}
\ee
which is a complex-valued function, unlike $Ai(x)$.
It can readily be seen that the amplitude of the field (\ref{evolvz}) is both shape invariant
and self-accelerating, since
\be
 |Ai(x,z)| = |Ai(x-z^2/4)| ,
\ee
i.e., the amplitude at the output plane $z$ is exactly the same as the amplitude resulting
from moving the amplitude at $z=0$ from $x$ to $x - z^2/4$:
Moreover, it is also seen that the whole field undergoes a net displacement that goes
with the square of $z$, which, in analogy to Airy wave packets, is associated with
an acceleration.

Consider now an arbitrary coherent superposition of Airy beams at the input plane $z=0$,
where each of these beams is affected by a random displacement $\lambda$ and contributes to
the superposition with an amplitude $c(\lambda)$.
This combination produces a field amplitude
\be
 \bar{\mathcal{U}}(x,0) = \int c(\lambda) Ai(x-\lambda) \ \! d\lambda .
 \label{eq7}
\ee
By virtue of the orthogonality relation (\ref{ortho}), we have
\be
 c(\lambda) = \int \bar{\mathcal{U}}(x,0) Ai(x-\lambda) \ \! dx .
 \label{eq8}
\ee
Because of the linearity of the superposition (\ref{eq7}), at an arbitrary plane $z$, the
field amplitude $\bar{\mathcal{U}}$ read as
\be
 \bar{\mathcal{U}}(x,z) = \int c(\lambda) Ai (x - \lambda, z) \ \! d\lambda .
 \label{eq7bis}
\ee
Appealing to the identity (\ref{evolvz}), this field amplitude can be recast as
\be
 \bar{\mathcal{U}}(x,z) = \int c(\lambda) e^{i(x - \lambda - z^2/6)z/2}
  Ai (x - \lambda - z^2/4) \ \! d\lambda .
 \label{eq11}
\ee
%
%

The field amplitude (\ref{eq11}) describes a fully coherent beam with analogous properties
to those displayed by the constituting Airy beams.
Let us consider the case of a number of random realizations of $\bar{\mathcal{U}}(x,z)$.
At $z=0$, the CSD describing this field reads as
\ba
 \mathcal{W}(x,x',0) & = &
 \langle \bar{\mathcal{U}}^*(x,0) \ \! \bar{\mathcal{U}}(x',0) \rangle \nonumber \\
 & = & \iint \mathcal{C}_0(\lambda,\lambda') Ai(x-\lambda) Ai(x'-\lambda') \ \! d\lambda d\lambda' , \nonumber \\
 \label{eq9}
\ea
where
\be
 \mathcal{C}_0(\lambda,\lambda') \equiv \langle c^*(\lambda) c(\lambda') \rangle
 \label{eq10}
\ee
is a non-negative Hermitian function.
This function provides us with information on possible
correlations between two different displacements $\lambda$ and $\lambda'$.
Because (\ref{eq9}) has the functional form of a convolution integral, its spectrum in
the Fourier plane will read as the product of a non-negative Hermitian function of the
corresponding momenta $k$ and $k'$, and the typical cubic phase factors that characterize
the spectrum of Airy beams.
At any other $z$-plane, the CSD reads as
\begin{widetext}
\ba
 \mathcal{W} (x,x',z) & = & e^{i(x' - x)z/2} \iint \mathcal{C}_0(\lambda,\lambda')
  Ai^*(x - \lambda,z) Ai(x' - \lambda',z) \ \! d\lambda d\lambda' \nonumber \\
 & = & \iint \mathcal{C}_0(\lambda,\lambda') e^{i(\lambda - \lambda')z/2}
  Ai(x - \lambda - z^2/4) Ai(x' - \lambda' - z^2/4) \ \! d\lambda d\lambda' ,
 \label{eq13}
\ea
\end{widetext}
which can be written in a more compact manner as
\be
 \mathcal{W}(x,x',z) = e^{i(x' - x)z/2} \mathcal{W}_0 (x,x',z) .
 \label{eq12}
\ee
In this latter expression, $\mathcal{W}_0$ denotes the result from the double integral
in Eq.~(\ref{eq13}), which is the quantity that will be examined to determine whether
shape-invariance and self-acceleration still remain.
Note that the complex exponential prefactor in Eq.~(\ref{eq12}) only adds a fast oscillation
that masks these behaviors [similar to the role played by the prefactor on the right-hand
side of (\ref{evolvz})].

The above CSD can be shown to render a family of genuine diffraction-free, partially
coherent sources, such that their intensity profile is shape-invariant with $z$, and the
absolute value of their degree of coherence and their which-path distinguishability both
exhibit transverse translation while propagating along $z$.
A sufficient condition for this to happen consists in choosing totally uncorrelated
displacements $\lambda$ and $\lambda'$, such that
\be
 \mathcal{C}_0(\lambda,\lambda') = P(\lambda) \delta(\lambda - \lambda') ,
 \label{eq16}
\ee
with $P(\lambda) > 0$ to ensure that $\mathcal{W}$ is a genuine CSD.
It can readily be noticed that this choice leads to
\be
 \mathcal{W} (x,x',z) = e^{i(x' - x)z/2} \ \! \mathcal{W}_0 (x - z^2/4, x' - z^2/4, 0) ,
 \label{eq17}
\ee
since
\ba
 \mathcal{W}_0 (x,x',z) & = &
  \int P(\lambda) Ai(x - \lambda - z^2/4) \nonumber \\ &  & \qquad \qquad
   \times Ai(x' - \lambda - z^2/4) \ \! d\lambda \nonumber \\
  & = & \mathcal{W}_0 (x -z^2/4,x'-z^2/4,0) .
 \label{eq15}
\ea
The latter expression shows that shape invariance and
self-acceleration are both guaranteed, as it was pointed out above.
Furthermore, the same properties also hold for the amplitude of the CSD (\ref{eq17}),
since
\be
 \left\arrowvert \mathcal{W}(x,x',z) \right\arrowvert =
  \left\arrowvert \mathcal{W} (x -z^2/4,x'-z^2/4,0) \right\arrowvert ,
 \label{eq14}
\ee
although it still describes an infinite energy beam, since the beam is spatially unbound,
like an ideal Airy beam.
This can readily be seen from the expression for the associated intensity distribution,
which reads as
\ba
 I(x,z) & = & I(x-z^2/4) \nonumber \\
 & = & \int P(\lambda) Ai(x-\lambda - z^2/4) Ai(x-\lambda - z^2/4) \ \! d\lambda .
 \nonumber \\ & &
 \label{eq18}
\ea
It is observed that, effectively, the integral of this quantity over $x$, at any $z$-plane, is
unbound.
Furthermore, it is worth noting how, from the flow associated with these partially coherent
Airy beams (with infinite energy), it is readily inferred the self-accelerating effect.
Specifically, the expression for the associated flow is
\ba
 {\bf j}(x,z) & = & c \omega^2 \Big\{
  i \left[ \frac{\partial \mathcal{W}(x,x',z)}{\partial x}
    - \frac{\mathcal{W}(x,x',z)}{\partial x'} \right]
   \Bigg\rvert_{x'=x} \hat{\bf x} \nonumber \\ &  &  \qquad \qquad
  + 2I(x,z) \hat{\bf z} \Big\} \nonumber \\
 & = & c \omega^2 \left( z \hat{\bf x} + 2 \hat{\bf z} \right) I(x - z^2/4) ,
 \label{eq19}
\ea
which is orthogonal to the curve $x-z^2/4$, with tangent vector
$\hat{\bf x} - z \hat{\bf z}/2$.
Accordingly, the flux described by Eq.~(\ref{eq19}) explicitly depends on the
profile displayed by the intensity distribution, regardless of the blurring that $P(\lambda)$
might induce on the oscillations on the left of the main maximum.
Therefore, the flux will describe an accelerated beam regardless of the choice of $P(\lambda)$.

Before assigning a particular functional form to $P(\lambda)$ and analyzing its
consequences, there are two situations of physical interest worth discussing.
On the one hand, if $P(\lambda)$ is a constant function, independent of $\lambda$, the
beams described by (\ref{eq17}) will be totally incoherent, since the averaging caused by
$P(\lambda)$ leads to a total suppression of the oscillations arising from by the
Airy functions.
On the other hand, if there is a high localization around a single $\lambda$ value, i.e.,
$P(\lambda)$ reduces to a Dirac delta function, $P(\lambda) \sim \delta(\lambda - \lambda_0)$,
the fully coherent Airy beam is recovered, but moved a distance equivalent to propagate the
beam from $z=0$ to $z = 2\sqrt{\lambda_0}$.
In both cases, though, the corresponding beam is still an infinite energy beam.
These two situations can be illustrated, without any loss of generality, with a Gaussian
spread function,
\be
 P(\lambda) = \frac{1}{\sqrt{\pi\sigma^2}}\ e^{-\lambda^2/\sigma^2}
  = \sqrt{\frac{\alpha}{\pi}}\ e^{-\alpha \lambda^2} ,
 \label{psf}
\ee
which smoothly approaches the above limits when $\sigma$ either goes to $\infty$
($\alpha \to 0$) or to 0 ($\alpha \to \infty$), respectively.

\begin{figure}[!t]
	\centering
	\includegraphics[width=\columnwidth]{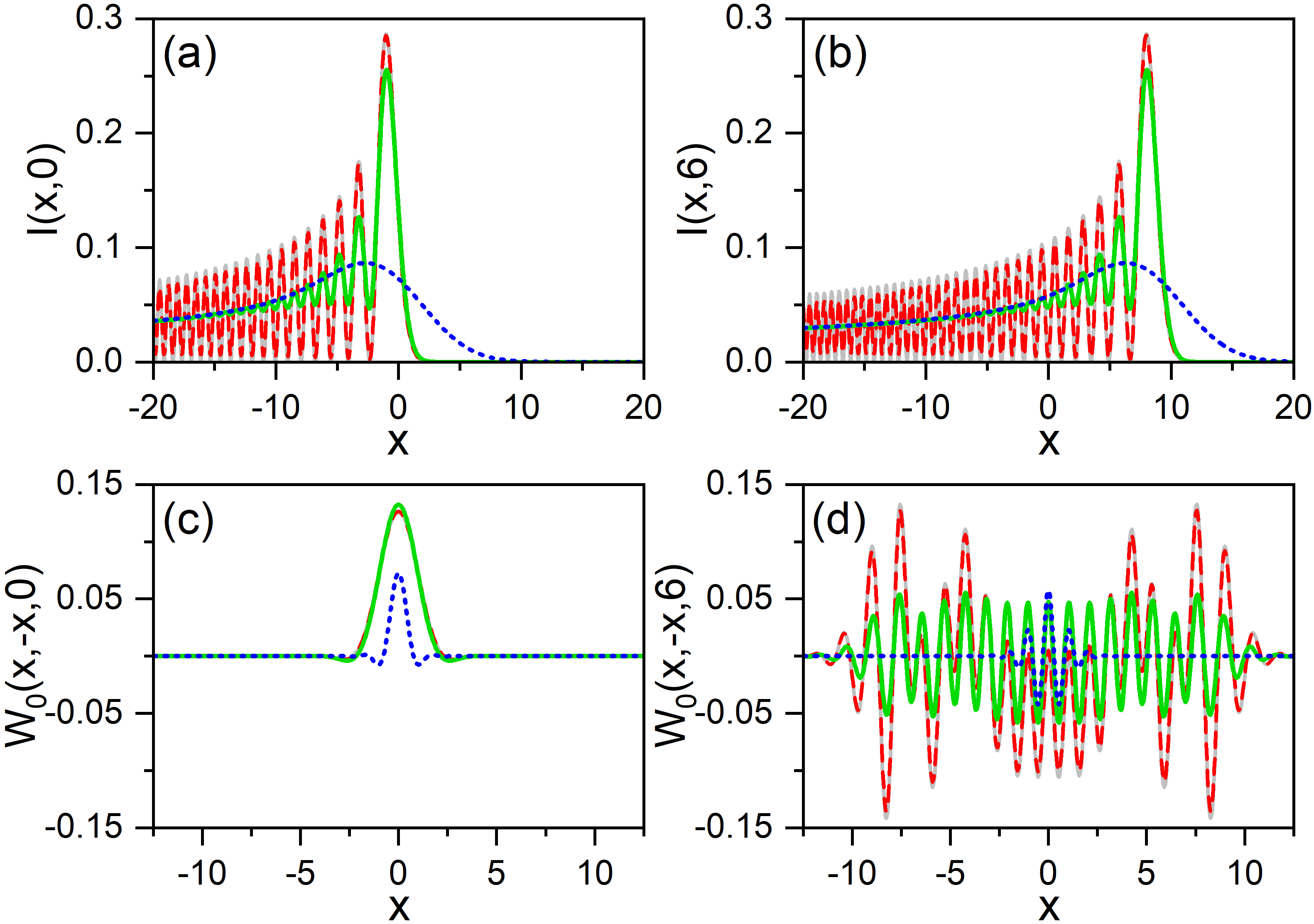}
	\caption{\label{fig1}
		Partially coherent Airy beam affected by the spread function $P(\lambda)$ (\ref{psf}).
		The intensity distribution is displayed in the top panels, while the amplitude $\mathcal{W}_0$
		is shown in the bottom ones for three different values of the spread parameter $\sigma$:
		$\sigma = 0.1$ (dashed red line), $\sigma = 0.5$ (solid green line), and $\sigma = 5$
		(dotted blue line).
		Both quantities have been computed at $z=0$ (left panels) and $z = 6$ (right panels) in
		order to make more apparent the shape-invariance and self-acceleration properties that
		characterize this type of infinite energy beams, as well as the gradual suppression of
		the spatial correlations (measured through $\mathcal{W}_0$) as $\sigma$ increases.
		To compare with, the fully coherent case (thin solid gray line), corresponding to
		$P(\lambda) \sim \delta(\lambda)$, has also been included in all cases.
        [Note that we are using dimensionless units (see text for details) and hence no units
        are specified either in this figure or the following ones.]}
\end{figure}

To better understand the action of (\ref{psf}) over the coherence of the infinite energy
Airy beam, consider the cases displayed in Fig.~\ref{fig1}, where the intensity distribution
is shown in the top panels [Figs.~\ref{fig1}(a) and \ref{fig1}(b)], while the amplitude $\mathcal{W}_0(x,-x,z)$,
with $x' = -x$, is represented in the bottom ones [Figs.~\ref{fig1}(c) and \ref{fig1}(d)].
Note that, to simplify the analysis, $\mathcal{W}_0(x,x',z)$ has been considered instead
of $\mathcal{W}(x,x',z)$, which avoids the effect of the fast oscillations due to the
the complex exponential prefactor in the latter.
The fully coherent Airy beam (i.e., the usual Airy beam) is represented in all cases with
the thin solid gray line.
As it can be seen, comparing Figs.~\ref{fig1}(a) and \ref{fig1}(b), for $z=0$ and $z=6$, respectively, the
beam is shape invariant, with all points of the latter having undergone a displacement
rightwards equivalent to $z^2/4 = 9$.
Note how the leading maximum at $x_{\rm max} \approx -1$ [see Fig.~\ref{fig1}(a)]
has moved to $x_{\rm max} \approx 8$ [see Fig.~\ref{fig1}(b)], which makes evident the
characteristic self-acceleration, although the full-width at half maximum remains the
same in both cases (${\rm FWHM} \approx 1.64$).
However, regarding the space correlations of the beam, described by $\mathcal{W}_0$, the
lower panels show a remarkable increase along the secondary diagonal ($x' = -x$), determined
by the development of long-range, fast oscillations.
This can be understood as an effect of the spatial overlapping of larger and larger portions
of an Airy beam traveling in one direction (say, positive $x$) with its mirror image,
traveling in the opposite direction (towards negative $x$).
Thus, at $z=0$, basically only the main lobes of both counter propagating beams overlap,
giving rise to a single (and positive) ``bump'', as seen in Fig.~\ref{fig1}(c).
However, at $z=6$, the effective overlapping covers, approximately, the region that goes
from $x \approx -10$ to $x \approx 10$, thus giving rise to the fast oscillations within
this range observed in Fig.~\ref{fig1}(d).

\begin{figure*}[!t]
	\centering
	\includegraphics[width=0.9\textwidth]{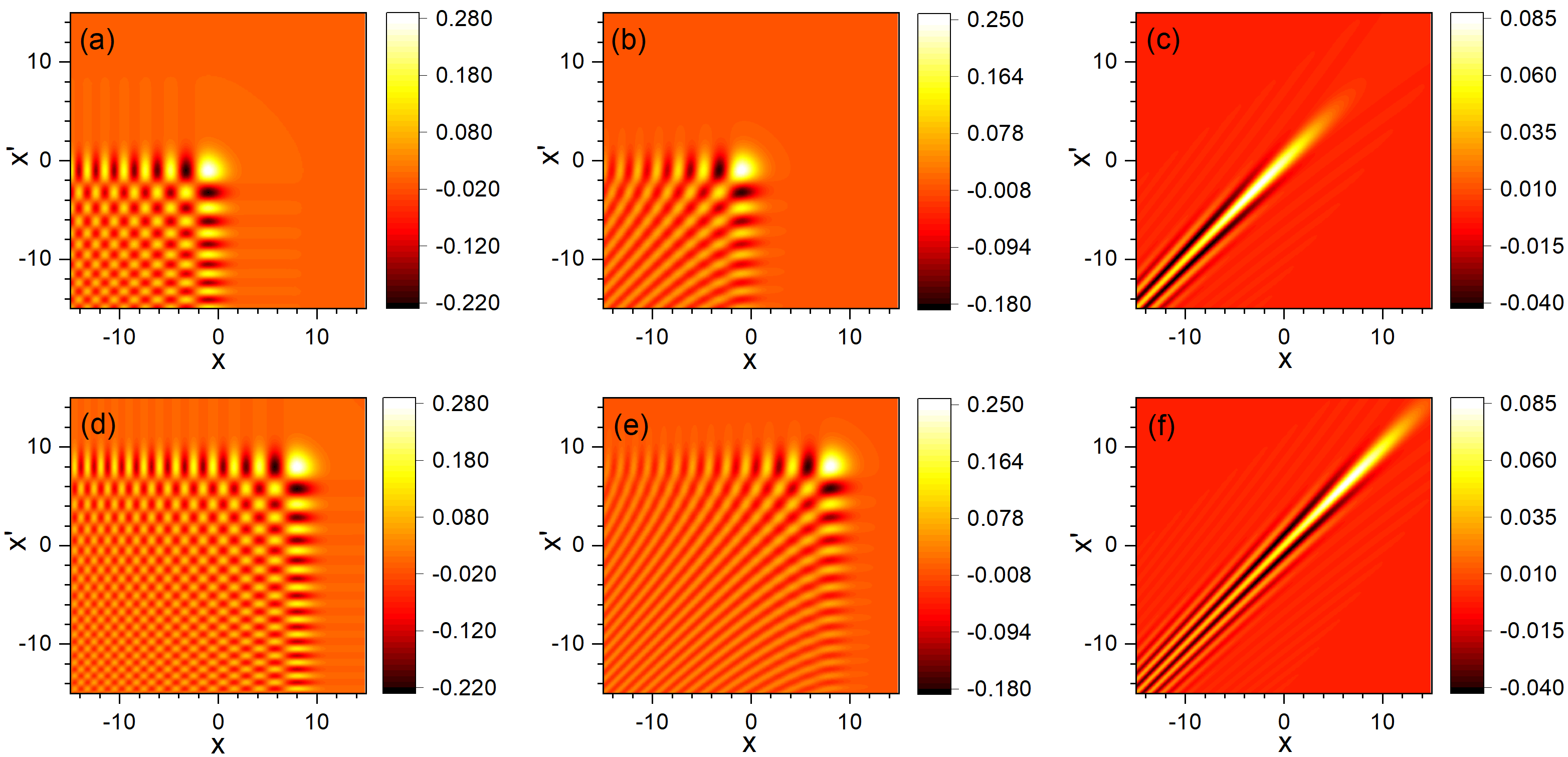}
	\caption{\label{fig2}
		Density plots describing the behavior of the amplitude $\mathcal{W}_0(x,x',z)$ for
		the three values of the spread parameter $\sigma$ considered above in Fig.~\ref{fig1}:
		$\sigma = 0.1$ (left column), $\sigma = 0.5$ (middle column), and $\sigma = 5$ (right column).
		To better perceive the shape invariance of the partially coherent beams and their
		self-acceleration, the same two cases for $z$ have also been considered: $z=0$
        (top row) and $z=6$ (bottom row).}
\end{figure*}

Let us now consider the effects of incoherence induced by (\ref{psf}).
Thus, if $\sigma$ is small compared to the FWHM associated with the main lobe, this maximum
as well as the nearby (on its left) maxima of the Airy beam will basically remain unaffected.
The effect will start being noticed as the width of the maxima that form the tail of the
beam become comparable with the value of $\sigma$.
This is what can be observed for $\sigma = 0.1$ (dashed red line in all panels in
Fig.~\ref{fig1}), particularly further away from the main lobe, as seen in Fig.~\ref{fig1}(b),
which translates into lower maxima and non-negligible minima.
Regarding $\mathcal{W}_0$, though, no important effects can be noticed either at $z=0$ [see
Fig.~\ref{fig1}(c)] or at $z=6$ [see panel Fig.~\ref{fig1}(d)], because within the $x$-range covered the Airy beam
does not experience important incoherence effects [see Figs.~\ref{fig1}(a) and \ref{fig1}(b)].
However, by further increasing the value of $\sigma$, to $\sigma = 0.5$ (solid green line),
the effect becomes more prominent, as seen in both top panels.
For this value of $\sigma$, only the leading maximum remains, while the faint signature
of few secondary maxima can also be perceived, approaching very quickly a decreasing tail.
Correspondingly, the oscillations displayed by $\mathcal{W}_0$ also undergo a remarkable damping.
Finally, in the regime of large $\sigma$, here illustrated with $\sigma=5$ (dotted blue line),
we find no traits of interference at all; the whole intensity distribution has been smoothed
out and now looks like the average value of the former Airy beam.
Essentially, coherence has almost totally been washed out; the only living trait can be
noticed through $\mathcal{W}_0$, with oscillations that remain pretty close to a small region
around $x=0$, as seen in Fig.~\ref{fig1}(d).
Nevertheless, in all these partially coherent cases, neither the shape-invariance property
nor the self-acceleration one disappear, but they are nicely preserved.

To better appreciate why in the large $\sigma$ range oscillations still persist in the CSD,
thus indicating the presence of spatial correlations in this regime, while the intensity
distribution looks like a distribution for a fully incoherence beam, in Fig.~\ref{fig2}
we have represented, in terms of density plots, the amplitude $\mathcal{W}_0$ for the
three values of $\sigma$ considered in Fig.~\ref{fig1} and the two values of $z$.
Thus, from top to bottom, $\sigma = 0.1$ (left column), $\sigma = 0.5$ (middle column), and
$\sigma = 5$ (right column); on the top row, the density plots represent the cases for
$z=0$, while on the bottom row, it is for $z=6$.
As it can be read in the color code legend, on the right of each panel, the transition from
negative to positive values is indicated with lighter and lighter shading.
Again, all cases shown clear evidence of the already mentioned shape-invariance and
self-acceleration properties regardless of $\sigma$, which indicates that these properties
are not characteristic traits of fully coherent Airy beams, but they can also be observed
in partially coherent Airy beams with infinite energy.
Furthermore, these density plots show that the effect of (\ref{psf}) consists of
annihilating the interferential (oscillatory) traits in a faster manner along the main
diagonal (which coincides with the intensity distribution) than along the secondary
diagonal, where there is a gradual ``collapse'' of $\mathcal{W}_0$ towards the main
diagonal as $\sigma$ increases.
This collapsing towards the main diagonal is, indeed, analogous to the effect that
decoherence has on the density matrix that specifies the state of a quantum system, when
the latter is influenced by a Markovian, Brownian-type environment \cite{sanz:CJC:2014}.

Finally, it is also worth noting that the above
family of CSDs can be further enlarged by adding a function $F(x'-x)$ to the
CSD~(\ref{eq13}) \cite{ponomarenko:OptLett:2021}, i.e.,
\ba
 \mathcal{W}_F (x,x',z) & = & \mathcal{W} (x,x',z) + F(x' -x) \nonumber \\
 & = & e^{i(x' - x)z/2} \ \! \mathcal{W} (x - z^2/4, x' - z^2/4, 0) \nonumber \\ &  & \qquad
  + F(x' -x) ,
 \label{eq20}
\ea
such that
\be
 F(x' - x) = \int \tilde{\mathcal{F}}(\eta) e^{i(x' - x)\eta} \ \! d\eta ,
 \label{eq21}
\ee
with
\ba
 \tilde{\mathcal{F}}(\eta) & > & 0 , \\
 \int \tilde{\mathcal{F}}(\eta) \ \! d\eta & \prec & \infty ,
\ea
With this gauge-type transformation, not only $\mathcal{W}_F(x,x',z)$ is well-defined
and satisfies the paraxial wave equation, but the intensity still remains shape-invariant.
This can easily be shown as follows.
From (\ref{eq20}), we have
\be
 I_F(x,z) = I(x,z) + F(0) .
\ee
%
where $F(0)$ is a constant with finite value, as it follows from the above properties
for $\tilde{\mathcal{F}}(\eta)$.
Thus, because $I(x,z)$ is shape-invariant, we find
\be
 I_F(x,z) = I_F(x - z^2/4,0) .
 \label{eq22}
\ee
On the contrary, the modulus of the complex degree of coherence (and hence other
coherence-based quantities, such as the visibility) will be sensitive to this
transformation.
As it can be noted from (\ref{eq20}), we have
\begin{widetext}
\ba
 |\mathcal{W}_F(x,x',z)|^2 & = & |\mathcal{W}(x,x',z)|^2 + |F(x'-x)|^2
  + 2 {\rm Re} \left[ F^*(x'-x) \mathcal{W}(x,x',z) \right] \nonumber \\
   & = & |\mathcal{W}(x - z^2/4, x' - z^2/4, 0)|^2 + |F(x'-x)|^2 \nonumber \\
  & & + 2 {\rm Re} \left[ e^{i(x' - x)z/2} F^*(x'-x) \mathcal{W}(x - z^2/4, x' - z^2/4, 0) \right] .
\ea
On the other hand, we also have
\ba
 |\mathcal{W}_F(x - z^2/4, x' - z^2/4, 0)|^2 & = & |\mathcal{W}(x - z^2/4, x' - z^2/4, 0)|^2 + |F(x'-x)|^2
  \nonumber \\ & &
  + 2 {\rm Re} \left[ F^*(x'-x) \mathcal{W}(x - z^2/4, x' - z^2/4, 0) \right] .
\ea
Comparing both expressions, we reach the following result:
\ba
 |\mathcal{W}_F(x,x',z)|^2 - |\mathcal{W}_F(x - z^2/4, x' - z^2/4, 0)|^2 & = &
 2 {\rm Re} \left\{ \left[ e^{i(x' - x)z/2} - 1 \right] F^*(x'-x) \mathcal{W}(x - z^2/4, x' - z^2/4, 0) \right\}
  \nonumber \\
 & = & 4 \sin \left[\frac{(x'-x)z}{4}\right] \nonumber \\ & &
  \quad \times {\rm Re}
    \left[ i e^{i(x' - x)z/4} F^*(x'-x) \mathcal{W}(x - z^2/4, x' - z^2/4, 0) \right] .
\ea
Taking into account that $\mathcal{W}(x - z^2/4, x' - z^2/4, 0)$ is a real quantity, the above
expression can be recast as
\ba
 |\mathcal{W}_F(x,x',z)|^2 - |\mathcal{W}_F(x - z^2/4, x' - z^2/4, 0)|^2 & = &
 4 \sin \left[\frac{(x'-x)z}{4}\right] \nonumber \\ & & \quad
 \times \mathcal{W}(x - z^2/4, x' - z^2/4, 0) \ \!
  {\rm Im} \left[ e^{-i(x' - x)z/4} F(x'-x) \right] , \nonumber \\ & &
\ea
\end{widetext}
which shows that, in general, the shape-invariance is not preserved for the complex degree
of coherence (\ref{eq4}) for this family of CSDs, since
\be
 |\gamma_F(x,x',z)| \neq |\gamma_F(x - z^2/4, x' - z^2/4,0)| .
 \label{eq23}
\ee
Therefore, concerning the preservation of quantities such as the modulus of the degree
of coherence or the which-path distinguishability, the use of infinite-energy partially
coherence beams described by the CSD (\ref{eq13}) is mandatory.


\section{Finite Energy Beams}
\label{sec3}

The partially coherent beams introduced in Sec.~\ref{sec2} correspond to ideal situations
that are no experimentally accessible in the laboratory.
Let us then investigate the feasibility of finite-energy partially coherent beams that
nearly preserve all properties of the infinite-energy ones seen above.
In this regard, two types of partially coherent beams are introduced with finite energy
and hence experimentally feasible.
To this end, some specific correlations among the displacements acting on the random field
amplitude realizations are considered, thus extending the CSD functional form obtained in
Sec.~\ref{sec2}.

In order to determine to what extent the above mentioned finite-energy partially coherent
beams keep the shape invariance as they propagate along $z$, we introduce a measure
of the overlapping or projection of the corresponding partially coherent CSD at that
$z$-plane, $\mathcal{W}_{i0}(x,x',z)$ ($i= {\rm I}, {\rm II}$ labels the respective CSD
type) with the input CSD shifted along the transverse direction a distance $z^2/4$,
$\mathcal{W}_{i0}(x - z^2/4,x' - z^2/4,0)$,
\begin{widetext}
\be
 \varepsilon_i(z) = \frac{\displaystyle \left\arrowvert
  \iint \mathcal{W}_{i,0}^* (x - z^2/4, x' - z^2/4, 0)
    \mathcal{W}_{i,0} (x,x',z) \right\arrowvert^2 dx dx' }
 {\displaystyle \iint
 \left\arrowvert \mathcal{W}_{i,0} (x - z^2/4, x' - z^2/4, 0) \right\arrowvert^2 dx dx' \
 \iint \left\arrowvert \mathcal{W}_{i,0} (x,x',z) \right\arrowvert^2 dx dx'} ,
 \label{eq34}
\ee
\end{widetext}
which satisfies $0 \le \varepsilon_i(z) \le 1$.
If $\mathcal{W}_{i0}(x,x',z)$ remains relatively shape-invariant,
that is, resembling the shifted CSD, $\mathcal{W}_{i0}(x-z^2/4,x'-z^2/4,0)$, then
$\varepsilon_i(z)$ will be close to the unity, because there will be an important
overlapping between both amplitudes.
On the the contrary, if finite-energy effects are relevant, then the behavior of both CSDs will
divert very quickly, in relatively short distances $z$, and hence $\varepsilon_i(z)$ will exhibit a fast
falloff to zero with $z$.


\subsection{Type-I CSD}
\label{sec31}

The first type of beams are such that, at the input plane $z=0$, their CSD is given by
\be
 \mathcal{W}_{\rm I} (x,x',0) =
  \iint \mathcal{C}_I(\lambda,\lambda') Ai(x -\lambda) Ai(x' -\lambda')\ \! d\lambda d\lambda' ,
 \label{eq24}
\ee
in analogy to Eq.~(\ref{eq9}), with
\be
 \mathcal{C}_I(\lambda,\lambda') = \mathcal{P}(\lambda,\lambda') \mathcal{Q}(\lambda-\lambda') ,
\ee
which must be a non-negative definite Hermitian function.
The above two correlation functions (bivariate distributions) are required to satisfy the
conditions
\ba
 \mathcal{Q}(0) & \prec & \infty , \\
 \int \mathcal{P}(\lambda,\lambda) \ \! d\lambda & \prec & \infty ,
 \label{eq25}
\ea
for $\mathcal{W}_{\rm I}(x,x',0)$ to be well-defined and to carry finite energy.
Note that, in the particular case
\be
 \mathcal{Q}(\lambda - \lambda') \sim \delta(\lambda - \lambda') ,
\ee
the CSD (\ref{eq24}) will be close to the infinite energy CSD (\ref{eq9}), thus exhibiting
analogous properties to the latter.
As for the propagated form of Eq.~(\ref{eq24}), following the prescription given in
Sec.~\ref{sec2}, we find
\be
\mathcal{W}_{\rm I}(x,x',z) = e^{i(x' - x)z/2} \mathcal{W}_{{\rm I},0} (x,x',z) ,
\label{eq28}
\ee
with
\ba
 \mathcal{W}_{{\rm I},0} (x,x',z) & = &
  \iint \mathcal{C}_{\rm I}(\lambda,\lambda') e^{i(\lambda - \lambda')z/2} Ai(x - \lambda - z^2/4)
  \nonumber \\ &  & \qquad
   \times Ai(x' - \lambda' - z^2/4) \ \! d\lambda d\lambda' .
 \label{eq29}
\ea
Note that, due to the correlation between $\lambda$ and $\lambda'$, now it is not possible to
ensure that the propagated solution is either shape invariant or self-accelerating.

To further investigate this type of CSDs, let us now assign a functional form to the
bivariate distributions $\mathcal{P}$ and $\mathcal{Q}$ satisfying the above requirements
and, at the same time, that are analogous to functional forms that can be found in the
literature for CSDs similar to the one defined by Eq.~(\ref{eq24}) \cite{lumer:Optica:2015}.
Accordingly, consider
\ba
 \mathcal{P}(\lambda,\lambda') & = & e^{-\alpha(\lambda^2 + \lambda'^2)} ,
 \label{eq26} \\
 \mathcal{Q}(\lambda-\lambda') & = & e^{-\beta(\lambda - \lambda')^2} ,
 \label{eq27}
\ea
from which we obtain
\be
 \mathcal{C}_{\rm I}(\lambda,\lambda') = e^{-\alpha(\lambda^2 + \lambda'^2)}
  e^{-\beta(\lambda - \lambda')^2} .
 \label{CI}
\ee
Regarding $\mathcal{P}$, it accounts for the separate effect of the point spread functions
associated with the random displacements $\lambda$ and $\lambda'$, each given by a
Gaussian distribution, like (\ref{psf}), with $\alpha$ determining the spread range
or width of the distribution.
On the other hand, $\mathcal{Q}$ introduces the correlation between $\lambda$ and $\lambda'$,
controlled by means of a $\beta$ parameter, such that, as $\beta$ gets larger and larger,
$\mathcal{Q}$ approaches the aforementioned Dirac $\delta$ distribution.
For computational convenience, to establish a better comparison among different cases in the
results shown below, we are going to introduce a multiplicative normalizing prefactor
$\mathcal{N}_{\rm I}$ in $\mathcal{W}_{\rm I}(x,x',z)$, which arises from considering
\be
 \iint \mathcal{C}_{\rm I}(\lambda,\lambda') \ \! d\lambda d\lambda' = 1 .
\ee
This prefactor, which reads as
\be
 \mathcal{N}_{\rm I} = \frac{\sqrt{\alpha(\alpha + 2\beta)}}{\pi} ,
\ee
has thus been considered in the numerical calculations involving both $\mathcal{W}_{\rm I}$
and the associated intensity, $I_{\rm I}(x,z)$, shown and discussed below.

Taking into account the above expressions for $\mathcal{P}$ and $\mathcal{Q}$, let us now
investigate how the amplitude of the CSD (\ref{eq28}), $\mathcal{W}_{{\rm I},0}(x,x',z)$,
deviates from a shape-invariant and self-accelerating behavior, described by the amplitude
corresponding to moving a distance $z^2/4$ in both $x$ and $x'$ the amplitude of the CSD
(\ref{eq24}).
Thus, substituting the amplitudes of both (\ref{eq28}) and (\ref{eq24}) into Eq.~(\ref{eq34})
leads to
\be
 \varepsilon_{\rm I}(z) = \frac{\displaystyle \left\arrowvert
 \iint \left\arrowvert \mathcal{C}_{\rm I}(\lambda,\lambda')
  e^{i(\lambda-\lambda')z/2} \right\arrowvert^2 \ \! d\lambda d\lambda' \right\arrowvert^2}
 {\displaystyle \left[ \iint
 \left\arrowvert \mathcal{C}_{\rm I}(\lambda,\lambda') \ \! d\lambda d\lambda' \right\arrowvert^2 \right]^2 } .
 \label{eq35}
\ee
Substituting Eqs.~(\ref{eq26}) and (\ref{eq27}) into this expression leads to
\be
 \varepsilon_{\rm I}(z) = e^{-z^2/(8\alpha + 16\beta)} ,
 \label{eq40}
\ee
which depends on the parameters describing the spread of both distributions.
In other words, the correlation between displacements is, in principle, at the same level
in relevance as the independent spread functions, unless one of them clearly prevails over
the other (which can be done by properly tuning $\alpha$ and $\beta$).
Furthermore, also note that (\ref{eq40}) also introduces a scale along the $z$-direction
\be
 z_{\rm I} = \sqrt{8(\alpha + 2\beta)} ,
 \label{eq40b}
\ee
such that, when $z = z_{\rm I}$, the overlapping between the two amplitudes reduces to about
37\%.
This means that the fidelity between both amplitudes, and hence the preservation of the
two Airy-beam properties, will be guaranteed for $z \ll z_{\rm I}$, while the propagated
amplitude will lose these traits as $z$ increases (particularly, for $z \gg z_{\rm I}$).

\begin{figure}[!ht]
	\centering
	\includegraphics[width=\columnwidth]{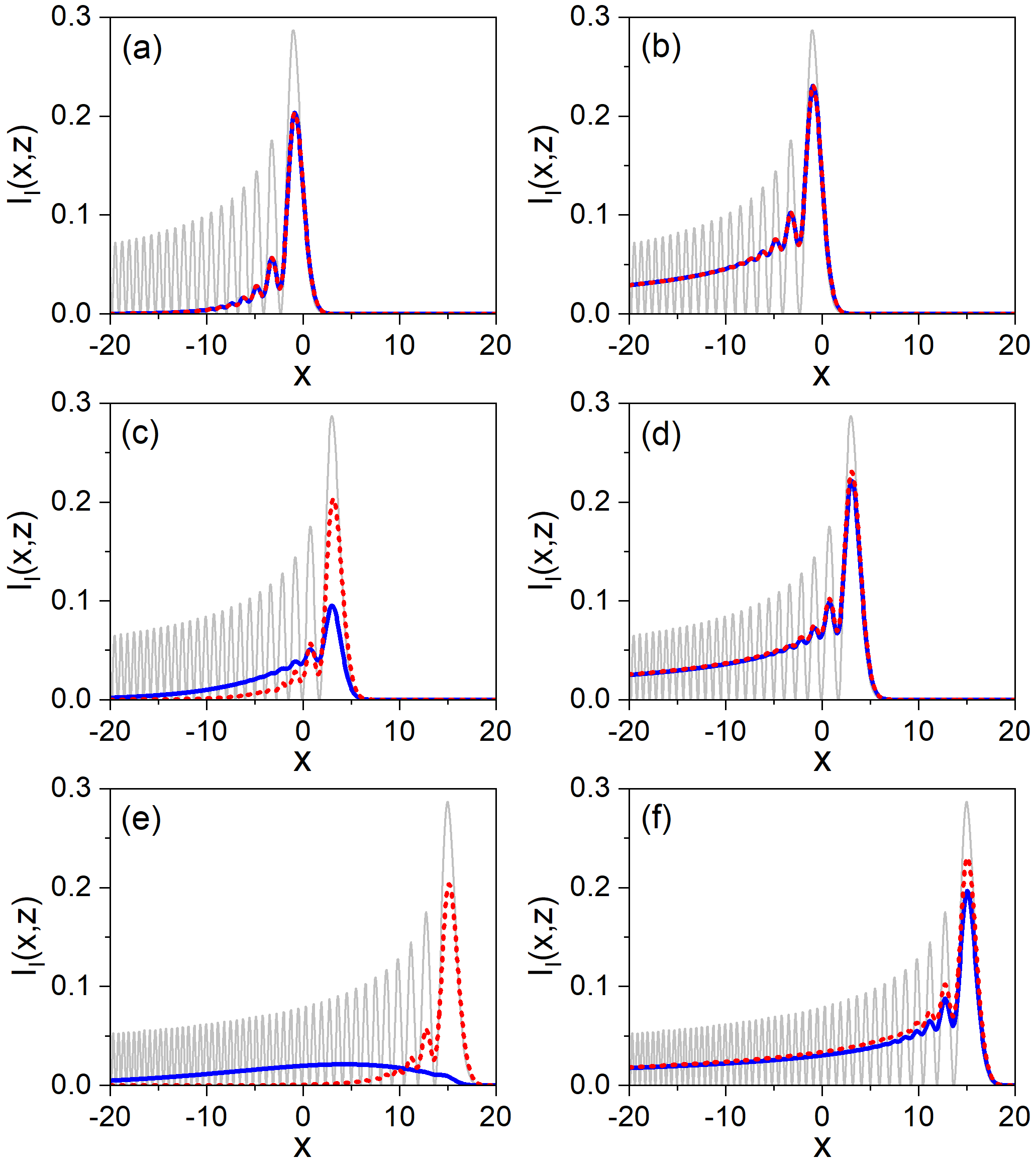}
	\caption{\label{fig3}
		Intensity distribution $I_{\rm I}(x,z)$ for $\alpha = 1$ and: $\beta = 0.5$
		(left column) and $\beta = 24.5$ (right column).
		In each panel, $I_{\rm I}(x,z)$ is denoted with the solid blue line, while the
        dotted red line represents $I_{\rm I}(x-z^2/4,0)$; to compare with, the intensity
        of the fully coherent Airy beam is also included (thin solid gray line).
		From top to bottom: $z = 0$, $z = 4$, and $z = 8$.}
\end{figure}

\begin{figure}[!t]
 \centering
 \includegraphics[width=\columnwidth]{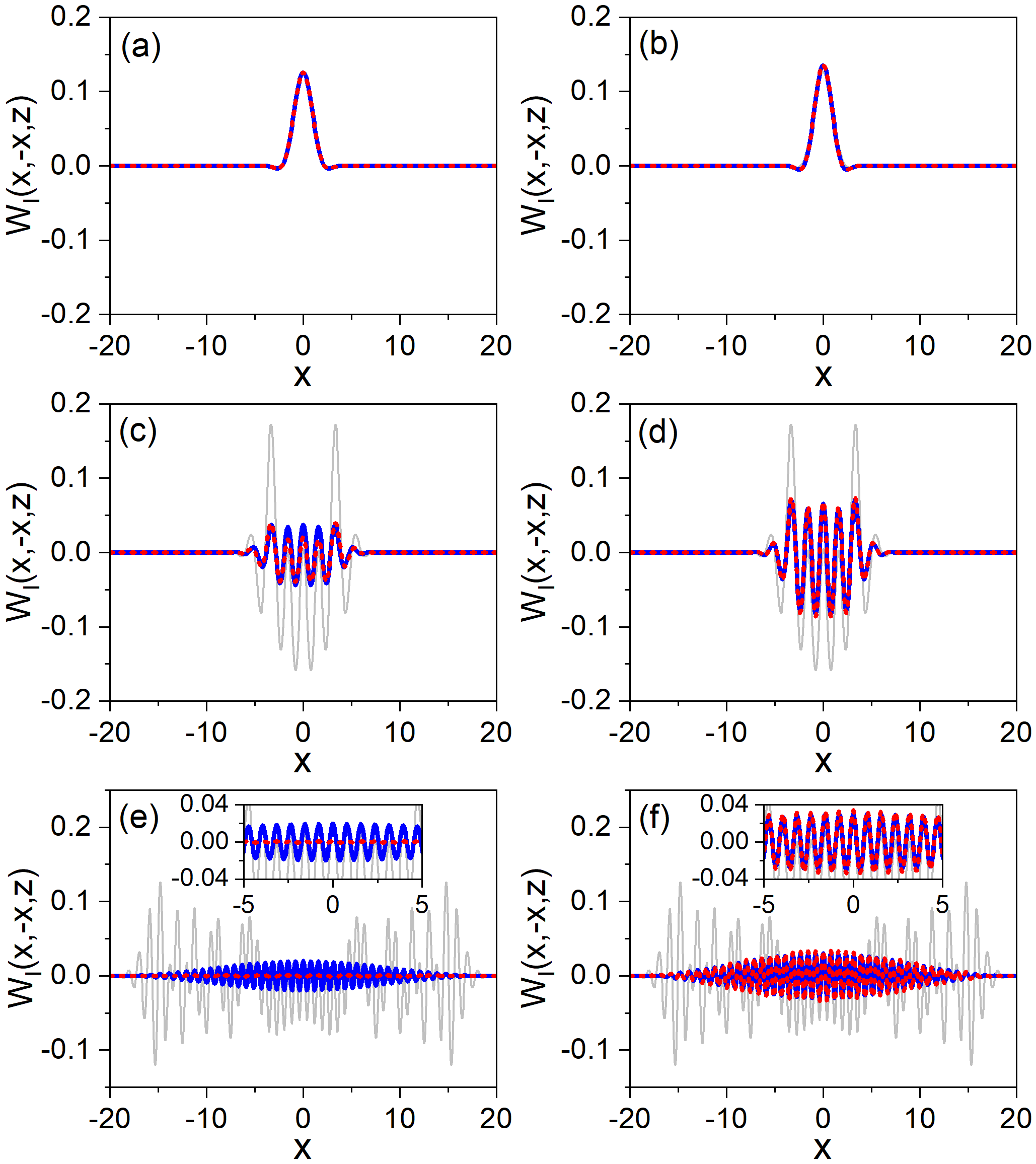}
 \caption{\label{fig4}
  Function $\mathcal{W}_{{\rm I},0}(x,-x,z)$ for $\alpha = 1$ and: $\beta = 0.5$ (left column)
  and $\beta = 24.5$ (right column).
  In each panel, $\mathcal{W}_{{\rm I},0}(x,-x,z)$ is denoted with the solid blue line, while
  the dotted red line represents $\mathcal{W}_{{\rm I},0}(x-z^2/4,-x-z^2/4,0)$; to compare
  with, the intensity of the fully coherent Airy beam is also included (thin solid gray line).
  From top to bottom: $z = 0$, $z = 4$, and $z = 8$.
  The insets in panels (e) and (f) provide an enlarged view of the central part of the amplitudes to
  better appreciate the deviations among them.}
\end{figure}

Figures~\ref{fig3} and \ref{fig4} show, respectively, the intensity and the amplitude
$\mathcal{W}_{{\rm I},0}$, with $x' = - x$, for different values of $z$ (increasing from
top to bottom) and two values of the parameter $\beta$.
In all graphs, the same value $\alpha = 1$ is considered, such that Eq.~(\ref{eq26})
corresponds to a Gaussian distribution of width $\sigma = 1$, i.e., in the intermediate
range of coherence, following the results discussed in Sec.~\ref{sec2}).
Regarding the values assigned to $\beta$, they have been chosen taking into account the
discussion about $z_{\rm I}$.
Specifically, we have considered $\beta = 0.5$ and 24.5, for which $z_{\rm I} = 4$ and
20, respectively.
If we compare the intensity distributions displayed in Figs.~\ref{fig3}(a) and \ref{fig3}(b),
for $z=0$, we find that the overall profile is essentially determined by $\alpha$ (similar damped
oscillatory behavior in both cases, compared to the fully coherent Airy beam, denoted with
the thin solid gray line), but the reach is longer in the case of larger $\beta$.
Accordingly, it is clear that, while $\mathcal{P}$ rules the partial coherence of the beam, the
distribution $\mathcal{Q}$ is going to determine the energy content.
On the other hand, as indicated above regarding the physical meaning of $z_{\rm I}$, by
inspecting Figs.~\ref{fig3}(c) and \ref{fig3}(d), for $z=4$, and Figs.~\ref{fig3}(e) and \ref{fig3}(f),
for $z=8$, we readily notice that the statement is correct, as the intensity for larger $\beta$ (i.e.,
$\beta = 24.5$) basically preserves the shape-invariance and self-accelerating properties for the
$z$-range considered, while the same does not happen for small $\beta$.
In the latter case, not only the coherence is remarkably lost at $z=2$, but the intensity distribution
(solid blue line) is smeared out all over the place as $z$ further increases, thus loosing all information
about the initial profile, which becomes unrecognizable compared to the simply displaced intensity
distribution (dotted red line).

Regarding the coherence properties exhibited by the CSD, let us now focus on Fig.~\ref{fig4}.
As it is shown for $\beta = 0.5$ (see left column, from top to bottom), the oscillatory
behavior of the CSD evaluated at $x'=-x$ (solid blue line) is relatively weak compared to
the fully coherent case (thin solid gray line), particularly at $z=8$ [see
Fig.~\ref{fig4}(f)].
Although the intensity profile is rather unstructured (it resembles an asymmetric Gaussian),
the fact that it spans a relatively long distance ensures that the overlapping integral
(\ref{eq29}) does not vanish for a similar range, which warrants the permanence of the
oscillatory behavior in distances comparable with those covered by the fully coherent case.
Note that, on the contrary, if the displaced distribution is considered (see dotted red
line), because of its faster falloff (and hence a robust overlapping over much shorter
distances), such oscillatory behavior has already disappeared for $z=8$, where we observe
a nearly flat CSD [see Fig.~\ref{fig4}(f)].
This behavior is in contrast with what happens in the case of larger $\beta$ (see right
column), where both CSDs, the propagated one (solid blue line) and the displaced one (dotted
red line), exhibit basically the same behavior, because the overlapping integral is nonzero
over a larger distance in the second case, as a consequence of the shorter range of
$\mathcal{Q}$.
Of course, there is still a gradual cancellation of the oscillations as $x$ increases
in both cases, but note that this arises from the smoothing produced by the
$\mathcal{P}$ distribution.


\subsection{Type-II CSD}
\label{sec32}

Concerning the second type of beam, we define its CSD as follows
\be
 \mathcal{W}_{\rm II} (x,x',0) =
  \int \mathcal{R}(\lambda'') S^*(x-\lambda'') S(x'-\lambda'')\ \! d\lambda'' ,
 \label{eq30}
\ee
which is well defined if $\mathcal{R}(\lambda) > 0$, and where
\be
 S(x) = \int \tilde{\mathcal{S}}(\lambda) Ai(x - \lambda)\ \! d\lambda
 \label{eq31}
\ee
describes a finite-energy Airy beam
\cite{ponomarenko:OptLett:2021,lumer:Optica:2015,christodoulides:OptLett:2007,forbes:NatCommun:2013,arie:OptLett:2011,liu:OptExpress:2020} if
\be
  \iint \mathcal{R}(\lambda'') |\tilde{\mathcal{S}}(\lambda - \lambda'')|^2 \ \! d\lambda d\lambda'' < \infty .
 \label{total}
\ee
Making explicit the substitution of (\ref{eq31}) into the CSD (\ref{eq30}), we obtain
\ba
 \mathcal{W}_{\rm II} (x,x',0) & = & \iiint
  \mathcal{R}(\lambda'')  \tilde{\mathcal{S}}^*(\mu) \tilde{\mathcal{S}}(\mu') Ai(x - \mu - \lambda'') \nonumber \\ &  & \qquad \quad
   \times Ai(x' - \mu' - \lambda'') \ \! d\mu d\mu' d\lambda'' \nonumber \\
  & = & \iint \mathcal{C}_{\rm II}(\lambda,\lambda')
   Ai(x - \lambda) Ai(x' - \lambda') \ \! d\lambda d\lambda' , \nonumber \\ & &
 \label{eq30b}
\ea
with
\be
 \mathcal{C}_{\rm II}(\lambda,\lambda'') =
  \int \mathcal{R}(\lambda'') \tilde{\mathcal{S}}^*(\lambda - \lambda'') \tilde{\mathcal{S}}(\lambda' - \lambda'') \ \!
   d\lambda'' ,
 \label{eq37}
\ee
in analogy to Eq.~(\ref{eq9}), and where we have made use of the identity
\be
 \int \tilde{\mathcal{S}}(\mu) Ai(x - \mu - \lambda'') \ \! d\mu
 = \int \tilde{\mathcal{S}}(\lambda - \lambda'') Ai(x - \lambda) \ \! d\lambda ,
 \label{eq31b}
\ee
on the right-hand side of the second equality in Eq.~(\ref{eq30b}), with $\lambda = \mu + \lambda''$
[a similar expression holds for the terms depending on $\mu'$ in (\ref{eq30b}), with
$\lambda' = \mu' + \lambda''$].
Despite the apparently complicated functional form displayed by (\ref{eq30b}), the fact that
it represents a finite energy beam can readily be seen by integrating over $x$ the associated intensity
distribution.
%
%
After integration, one obtains the total power described by the expression (\ref{total}), which
is bound (finite).
Actually, if a Gaussian distribution is chosen for $\tilde{\mathcal{S}}$, then the finite-energy Airy beam
introduced in Ref.~\cite{christodoulides:OptLett:2007} is recovered.

Following the same procedure as before, we find that the propagated form of the CSD
(\ref{eq30}) reads as
\be
 \mathcal{W}_{\rm II} (x,x',z) = e^{i(x' - x)z/2} \mathcal{W}_{{\rm II},0} (x,x',z) ,
 \label{eq32}
\ee
with
\ba
 \mathcal{W}_{{\rm II},0} (x,x',z)
  & = & \iint \mathcal{C}_{\rm II}(\lambda,\lambda') e^{i(\lambda - \lambda')z/2}
   Ai(x - \lambda - z^2/4) \nonumber \\ &  & \qquad
  \times Ai(x' - \lambda' - z^2/4) \ \! d\lambda d\lambda' .
 \label{eq33}
\ea
Equation (\ref{eq32}) thus paves an alternative way to build partially coherent Airy
beams with finite energy.

To further investigate the behavior of this type of CSD, let us now consider some
particular functional forms for both $\mathcal{R}$ and $\tilde{\mathcal{S}}$.
More specifically, as in the previous cases analyzed, we also consider Gaussian
distributions, i.e.,
\ba
 \mathcal{R}(u) & = & e^{- a u^2} ,
 \label{eq41} \\
 \tilde{\mathcal{S}}(v) & = & e^{- b v^2} .
 \label{eq42}
\ea
As before, also for computational convenience, to set a comparison between difference
cases below, we consider
\be
 \iint \mathcal{C}_{\rm II}(\lambda,\lambda') \ \! d\lambda d\lambda' = 1 ,
\ee
which renders the renormalization prefactor
\be
 \mathcal{N}_{\rm II} = \frac{b}{\pi} \sqrt{\frac{a}{a+2b}} .
\ee
It is worth mentioning, though, that with the choice (\ref{eq41}) and (\ref{eq42}),
$\mathcal{C}_{\rm II}$ acquires a similar functional form to Eq.~(\ref{CI}) for
$\mathcal{C}_{\rm I}$ in terms of the distributions (\ref{eq26}) and (\ref{eq27}),
namely,
\be
 \mathcal{C}_{\rm II}(\lambda,\lambda') = e^{- \alpha' (\lambda^2 + \lambda'^2)}
   e^{- \beta' (\lambda - \lambda')^2} ,
 \label{CII}
\ee
with
\ba
 \alpha' & = & \frac{ab}{a + 2b} , \\
 \beta' & = & \frac{b^2}{a+2b} .
\ea
However, as it can be noticed, unlike $\mathcal{C}_{\rm I}$, here the spread factors,
$\alpha'$ and $\beta'$, involve features from both (\ref{eq41}) and (\ref{eq42}).
In other words, although each Airy beam is independently affected by a spread function
$\tilde{\mathcal{S}}$, the effect of $\mathcal{R}$ is to set a correlation between $\lambda$ and
$\lambda'$.

Two limits can be clearly identified.
For $a \gg b$, we obtain $\alpha' \approx b$ and $\beta' \approx b^2/a \ll b$.
Physically, this means that the spread range for $\lambda''$ is rather short and
hence we are going to observe nice oscillations in the corresponding intensity
distribution, although with a limited spatial extension.
On the other hand, for $a \ll b$, we find $\alpha' \approx a/2$ and $\beta' \approx b/2$,
that is, the two exponential factors in $\mathcal{C}_{\rm II}$ describe independent
distributions, as in the case of type-I CSDs.
In this case, because $|\lambda - \lambda'|$ can be relatively large, the oscillations
typical of the Airy function are expected to be partially suppressed (depending on the
actual value of the parameters $a$ and $b$).

\begin{figure}[!t]
	\centering
	\includegraphics[width=\columnwidth]{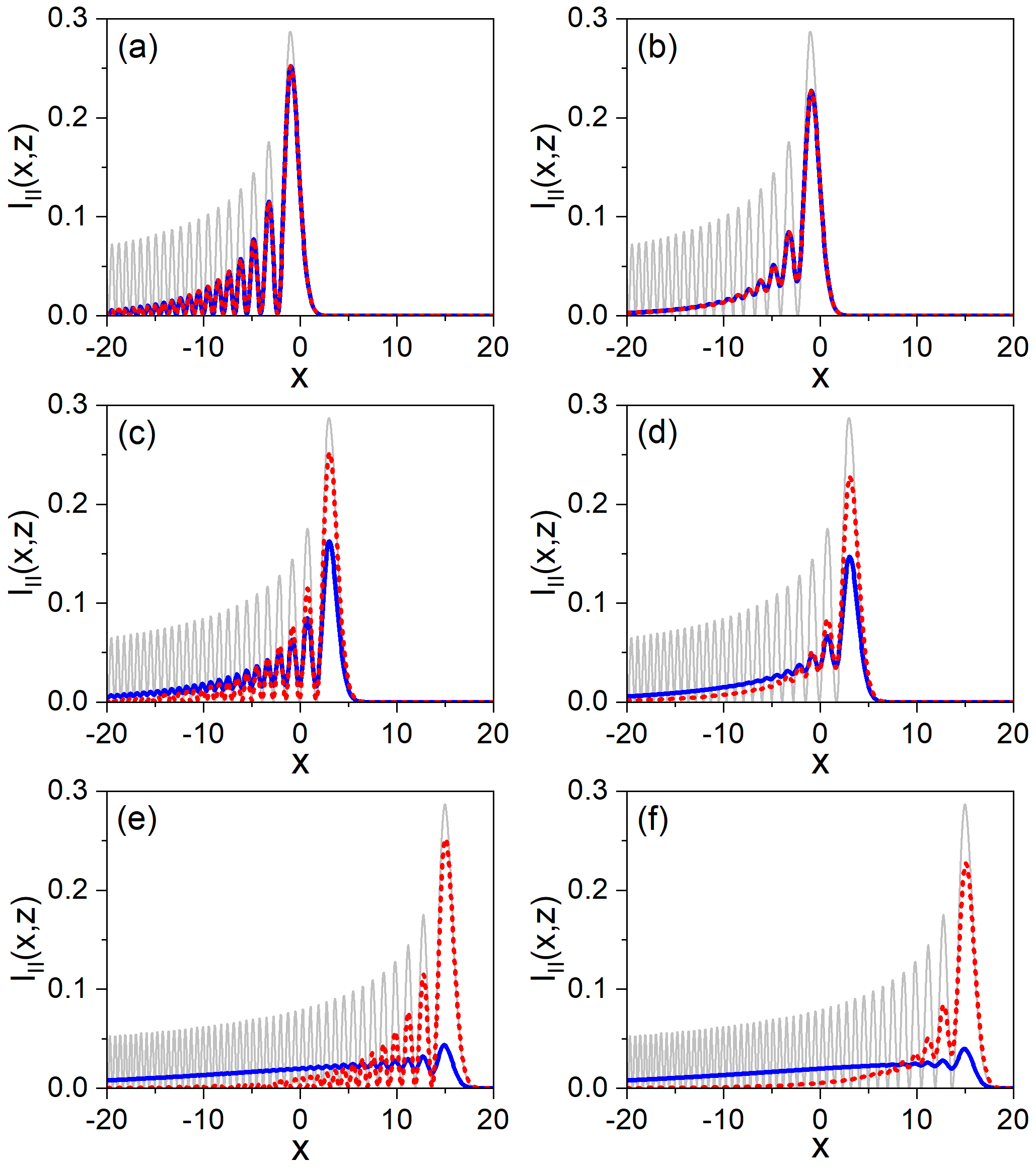}
	\caption{\label{fig5}
		Intensity distribution $I_{\rm II}(x,z)$ for $a = 4$ and: $b = 1$ (left column) and $b = 5$ (right column).
		In each panel, $I_{\rm II}(x,z)$ is denoted with the solid blue line, while the
        dotted red line represents $I_{\rm II}(x-z^2/4,0)$; to compare with, the intensity
        of the fully coherent Airy beam is also included (thin solid gray line).
		From top to bottom: $z = 0$, $z = 4$, and $z = 8$.}
\end{figure}

Here, the expression (\ref{eq34}) for the overlapping between the CSD (\ref{eq33})
and the CSD (\ref{eq30b}) propagated to the output plane $z$, reads as
\be
 \varepsilon_{\rm II}(z) =
  \frac{\displaystyle \left\arrowvert \iint \left\arrowvert
   \mathcal{C}_{\rm II}(\lambda,\lambda') \right\arrowvert^2
  e^{i(\lambda - \lambda')z/2} d\lambda d\lambda' \right\arrowvert^2}
 {\displaystyle \left[ \iint \left\arrowvert \mathcal{C}_{\rm II}(\lambda,\lambda')
  \right\arrowvert^2 d\lambda d\lambda' \right]^2} .
 \label{eq36}
\ee
After substituting (\ref{CII}) into this expression, we find
\be
 \varepsilon_{\rm II}(z) = e^{-z^2/16 b} .
 \label{eq43}
\ee
which only depends on $b$, unlike the correlation function $\mathcal{C}_{\rm II}$,
Eq.~(\ref{CII}), which depends on both parameters.
In this regard, each $b$ parameter determines a whole family of finite energy,
partially coherent beams, as all of them are going to behave the same way, although
with different degrees of incoherence, which is essentially governed by the
$a$ parameter.
Furthermore, each of these families obeys to the same decay length scale,
\be
 z_{\rm II} = 4 \sqrt{b} ,
\ee
i.e., regardless of the degree of coherence imposed by $a$, for all type-II beams the
degradation of the shape-invariance and self-acceleration properties are specified by
the same value of $z_{\rm II}$.

\begin{figure}[!t]
 \centering
 \includegraphics[width=\columnwidth]{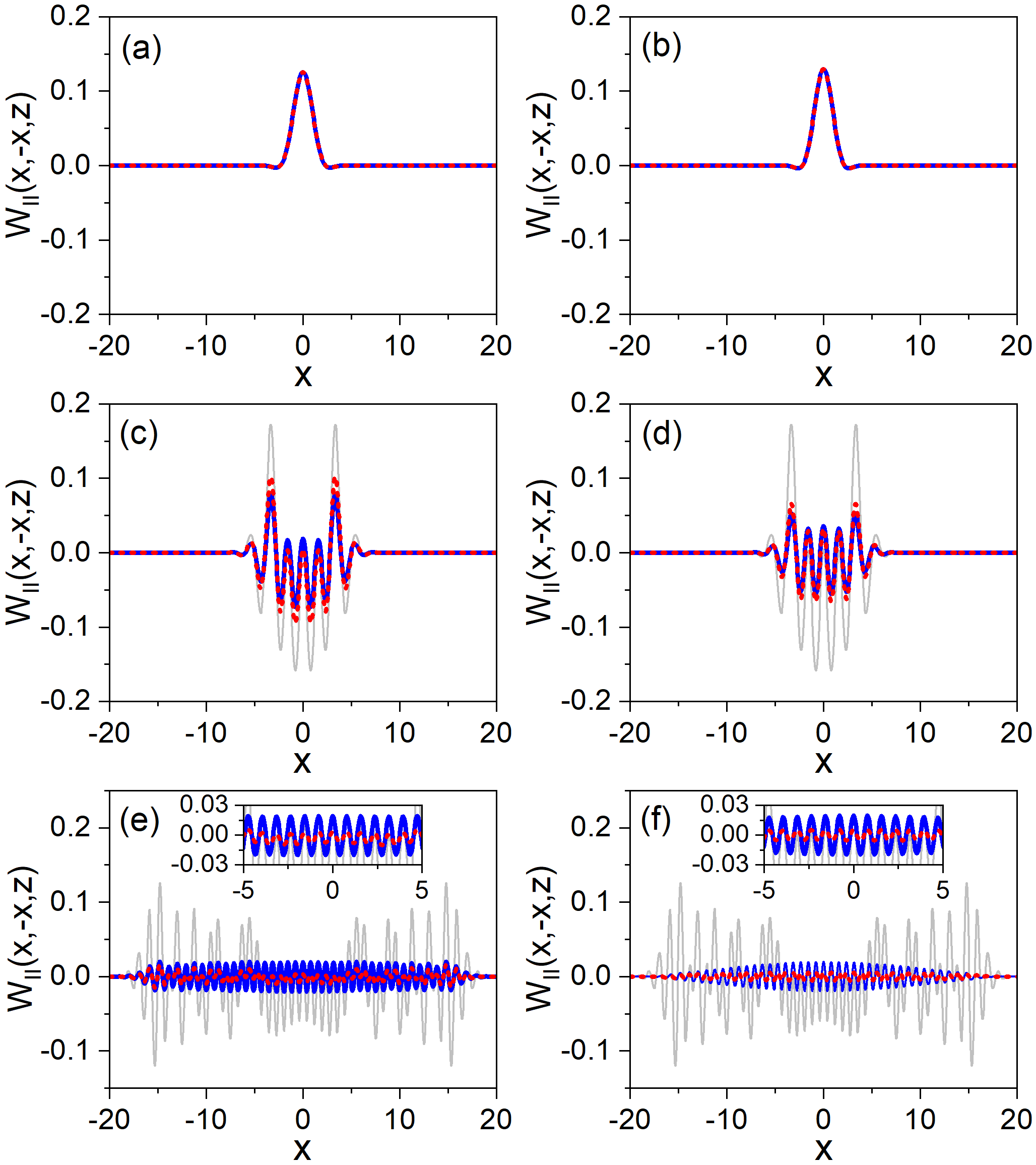}
 \caption{\label{fig6}
	Function $\mathcal{W}_{{\rm II},0}(x,-x,z)$ for $a = 4$ and: $b = 1$ (left column) and $b = 5$ (right column).
  In each panel, $\mathcal{W}_{{\rm II},0}(x,-x,z)$ is denoted with the solid blue line, while
  the dotted red line represents $\mathcal{W}_{{\rm II},0}(x-z^2/4,-x-z^2/4,0)$; to compare
  with, the intensity of the fully coherent Airy beam is also included (thin solid gray line).
	From top to bottom: $z = 0$, $z = 4$, and $z = 8$.
	The insets in (e) and (f) provide with an enlarged view of the central part of
	$\mathcal{W}_{{\rm II},0}(x,-x,z)$ and $\mathcal{W}_{{\rm II},0}(x-z^2/4,-x-z^2/4,0)$
	to better appreciate the deviations between them.}
\end{figure}

To better understand the above statement, in the intensity distribution for two type-II
beams with the same $b$ but different values of $a$ and at different $z$-planes (with
$z$ increasing from top to bottom) is shown in Fig.~\ref{fig5}.
In particular, we have chosen $b=4$, for which $z_{\rm II} = 8$.
Accordingly, the intensity distributions have been determined at $z=0$ (top), $z=4$
(middle), and $z=8$ (bottom) for $a=100$ (left column) and $a=4$ (right column), these two
values being in compliance with the limiting situations discussed above.
In particular, as seen in Figs.~\ref{fig5}(a) and \ref{fig5}(b), $a=100$ represents a
situation of high coherence, while $a=4$ describes a scenario with an important
suppression of coherence, analogous to the cases analyzed in Sec.~\ref{sec31}.
Moreover, also as in the previously analyzed case, the exponential-type decay tail, and
hence the energy content of the beam, is determined by $b$.
In order to get a better idea on the decrease of both coherence and energy, compare the
results for the two cases, $a=100$ and $a=4$, with the fully coherent Airy beam
(thin solid gray line).
Now, as $z$ increases, it can also be seen that the overlapping between the type-II beams
and the corresponding initial amplitudes moved rightwards by a quantity $z^2/4$ (dotted
red line) decreases in a similar manner, in compliance with (\ref{eq43}).
In Figs.~\ref{fig5}(c) and \ref{fig5}(d), for $z = 4$ ($= z_{\rm II}/2$), the correspondence
between both amplitudes falls to about 78\%, so both shape invariance and self-acceleration
can still be properly seen (although some deviations along the decaying tail can also be
observed).
However, in Figs.~\ref{fig5}(e) and \ref{fig5}(f), for $z=8$ ($= z_{\rm II}$), such
correspondence is much weaker, since the overlapping has fallen to about 37\%.
In this case, for both values of $a$, we can see that the beam has got a nearly homogeneous
distribution all over the place.

In Fig.~\ref{fig6} we have represented the amplitude $\mathcal{W}_{{\rm II},0}$ along the
secondary diagonal ($x'=-x$).
The results are pretty similar for both cases, with a
rather homogeneous oscillatory structure within the region covered by the beam.
Of course, in the case of shorter $a$, there is a faster decay towards the borders of
$\mathcal{W}_{{\rm II},0}$, as seen above for type-II CSDs.


\section{Concluding remarks}
\label{sec4}

Over recent years, different methods have been considered in the literature to generate
finite-energy coherent beams, which recreate the behavior of ideal Airy beams to a great
extent.
Some methods are based on the truncation of the angular momentum, which also induce
incoherence by removing and/or acting on some of the beam spectral components.
This has thus redirected the attention towards the design and experimental implementation of
partially coherent Airy beams, because of their intrinsic interest as a new resource of
structured light with important potential applications.
Unlike previous approaches considered in the literature, here we have focused on exploiting
the properties of the cross-spectral density (CSD) as an alternative strategy.
Thus, we have analyzed its behavior when incoherence is added by assuming an action from
external random fluctuations at the input plane.
In this regard, it has been seen that the shape-invariance and self-acceleration
properties displayed by fully coherent Airy beams can still be preserved in partially
coherent Airy beams provided they carry infinite energy and the fluctuations that
affect the field amplitude are totally uncorrelated.

From that starting point, by adding a certain correlation between the random fluctuations,
we have been able to design two types of partially coherent beams with finite energy, which
may display both shape-invariance and self-acceleration within a given range of the
propagation distance, $z$, before such properties are totally lost.
This $z$ range has been shown to depend on the parameter that specifies the distribution of
fluctuations.
Actually, this has been possible by defining a position-dependent parameter that
quantifies the degree of overlapping between the propagated beam and the input one displaced
by an amount equivalent to the propagation distance.
As it has been shown, depending on how the correlations between random displacements are
established, two types of CSDs have been introduced, for which this overlapping parameter
exhibits a Gaussian-type decay.
In what we have denoted as type-I CSD, the correlation among displacements is set on a local
level, i.e., directly applied between two any different Airy beams.
In this case, it has been seen that the decay rate depends on both the spread of the
random displacements and the spread range of the correlations between two any random
displacements.
However, if the correlations are set on a nonlocal level, i.e., they affect a whole swarm
of Airy beams that contribute to $S(x)$, then it has been shown that the decay
rate only depends on the spatial extent of the spread range of the spectral correlation
function $\tilde{\mathcal{S}}$.
This is the case for what we have denoted as type-II CSDs.
Note that, although in both cases one can determine how far the beam can be propagated with
its Airy-type properties being mostly preserved, the simpler dependence of type-II CSDs on
a single parameter enables the generation of an infinite family of partially coherent
Airy-type beams all behaving the same way regarding shape invariance and self-acceleration.

Finally, it is also worth mentioning that all the theory here developed can
be applied to the field of structured light beams as well as to the field of
matter waves and wave-packet design.
In this case, not only conditions for the design of particular particle beams
are provided, but such conditions can also be used to established the extent
of external factors that may act on the beams, leading to a suppression of
their coherence properties (induced either by incoherence or by decoherence).


\section*{Acknowledgments}

Financial support from the Spanish Agencia Estatal de
Investigaci\'on and the European Regional Development Fund (Project
PID2019-104268GBC21/AEI/10.13039/501100011033) is acknowledged.




%

\end{document}